\documentclass[epj]{webofc}
\usepackage[varg]{txfonts}   
\usepackage{textcomp}
\woctitle{3$^{\rm{rd}}$ International Conference on New Frontiers in Physics}
\begin{document}
\title{Light (Hyper-)Nuclei production at the LHC measured with ALICE}
%
%


\author{Francesco Barile for the ALICE Collaboration\inst{1}\fnsep\thanks{\email{francesco.barile@cern.ch}}}
\institute{Universit\`a degli Studi di Bari and INFN Bari}

\abstract{%
 The high center-of-mass energies delivered by the LHC during the last three years of operation led to accumulate a 
significant statistics of light (hyper-)nuclei in pp, p--Pb and Pb--Pb collisions.
The ALICE apparatus allows for the detection of these rarely produced particles over a wide
momentum range thanks to its excellent vertexing, tracking and particle identification capabilities. The last is based on
the specific energy loss in the Time Projection Chamber and the velocity
measurement with the Time-Of-Flight detector. The Cherenkov technique, exploited by a small
acceptance detector (HMPID), has also been used for the most central Pb--Pb collisions to identify
(anti-)deuterons at intermediate transverse momentum.\\
Results on the production of stable nuclei and anti-nuclei in pp, p--Pb and Pb--Pb collisions are presented.
Hypernuclei production rates in Pb--Pb are also described, together with a measurement of the
hypertriton lifetime. The results are compared with the predictions from thermal and coalescence
models. Moreover the results on the search for weakly-decaying light exotic states, such as the
$\Lambda\Lambda$ (H-dibaryon) and the $\Lambda$-neutron bound states are discussed.}
\maketitle
\section{Introduction}
\label{intro1}

Collisions of ultra-relativistic heavy ions provide a unique experimental condition to produce nuclei and hypernuclei thanks to the huge amount of energy deposited into a volume much larger than in pp collisions. 
The measurements presented here, have been performed in Pb--Pb collisions at $\sqrt{s_{\rm{NN}}}$~=~2.76~TeV as a function of collision centrality and in p--Pb collisions at $\sqrt{s_{\rm{NN}}}$~=~5.02~TeV as a function of charged-particle multiplicity. 

The unique particle identification capabilities of the ALICE detector~\cite{ALICE1} system is suited to measure nuclei and hypernuclei and for the search of exotic states like $\Lambda$n bound states and the H-dibaryon.
The production mechanisms of these particles are typically discussed within two approaches: the statistical thermal model and the coalescence model. In the thermal model~\cite{INTRO1, INTRO2, INTRO3} the chemical freeze-out temperature $T_{\rm{chem}}$
acts as the key parameter at LHC energies. The strong sensitivity of the nuclei production to the choice of $T_{\rm{chem}}$ is caused by the large mass $m$ and the exponential dependence of the yield given by the factor exp(-$m$/$T_{\rm{chem}}$). In the coalescence model, nuclei are formed by protons and neutrons which are nearby in space and exhibit similar velocities~\cite{INTRO4, INTRO5}. A quantitative description of this process, applied to many collision systems at various energies~\cite{INTRO6, INTRO7, INTRO9, INTRO10, INTRO11, INTRO12, INTRO13, INTRO14}, is typically based on the coalescence parameter $B_{\rm{A}}$ (see Section \ref{Nuclei1}). 
The two mechanisms give very similar predictions~\cite{INTRO15}.\\

\section{Particle identification in ALICE}
\label{Sec2}
In this section the particle identification (PID) detectors relevant for the analysis presented in this contribution are briefly described. A detailed review of the ALICE experiment and its PID capabilities can be found in~\cite{ALICE1, ALICE2, ALICE3}. The Inner Tracking System (ITS) is a six-layered silicon detector with radii ranging between 3.9 cm and 43 cm. The precise space-point resolution in the silicon layers allows a precise separation of primary and secondary particles (better than 300~$\mu$m) in the high track density region close to the primary interaction vertex. Four out of the six layers measure the particle specific energy loss per unit length (d$\textit{E}$/d$\textit{x}$) and are used for particle identification in the non-relativistic region. The Time Projection Chamber (TPC)  is the main central-barrel tracking detector of ALICE. It is a large-volume high-granularity cylindrical detector that provides three-dimensional hit information and d$\textit{E}$/d$\textit{x}$ measurement with up to 159 samples. It can measure charged-particle abundances on a statistical basis also in the relativistic rise for momenta up to 50 GeV/$c$. 
Figure~\ref{PERF} (left) shows the specific energy loss (d$E$/d$x$) in the TPC vs. rigidity ($p$/$z$) in Pb--Pb collisions at $\sqrt{s_{\rm{NN}}}~$~=~2.76~TeV. 
The Time Of Flight (TOF) detector is a large area array devoted to particle identification in the intermediate momentum range using the information on the velocity of the particle to identify them within an acceptance similar to that of the TPC. The total time resolution for tracks in Pb--Pb collisions is about 80 ps. 
The High Momentum Particle Identification Detector (HMPID)~\cite{highptpaper, COMOconf, MYPHD} is designed as a single-arm proximity-focusing Ring Imaging Cherenkov detector (RICH) where the radiator is a 15 mm thick layer of C$_{6}$F$_{14}$ (perfluorohexane) with a refraction index of $n$~=~1.2989 at the photon wave length $\lambda$~=~175~nm corresponding thus to the minimum particle velocity $\beta_{\rm{min}}$~=~0.77. The HMPID, located at about 5~m from the beam axis, consists of seven identical counters covering in total an acceptance of about 5$\%$ at central rapidity. 
Figure~\ref{PERF} (right) shows the squared particle masses calculated from the momentum and Cherenkov angle reconstructed by the ITS-TPC and HMPID, respectively, for Pb--Pb collisions at $\sqrt{s_{\rm{NN}}}$~=~2.76~TeV, for the 10$\%$ most central collisions.

\section{Nuclei}
\label{Nuclei1}

\begin{figure}
\centering
\includegraphics[width=6cm,clip]{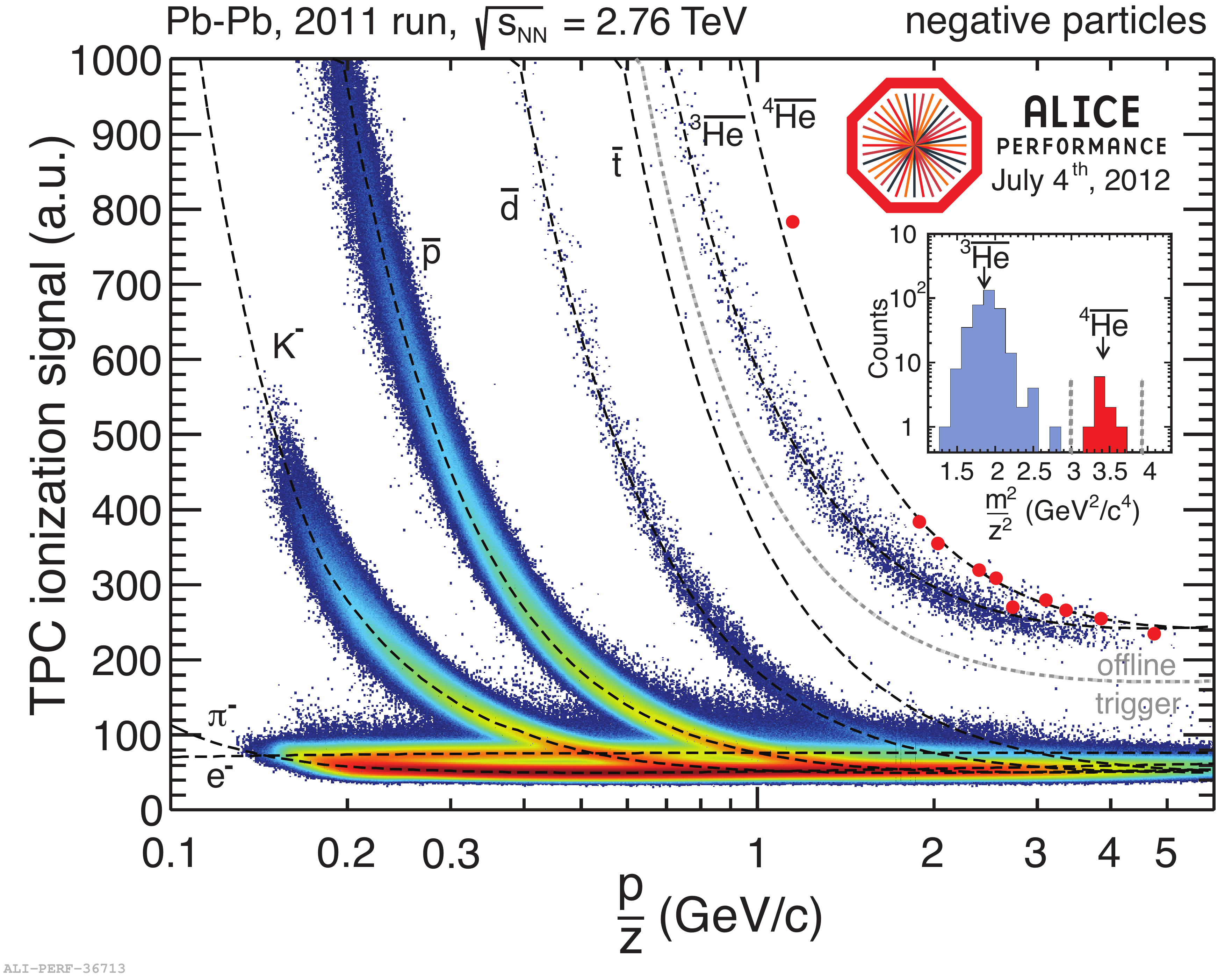}
\includegraphics[width=7cm,clip]{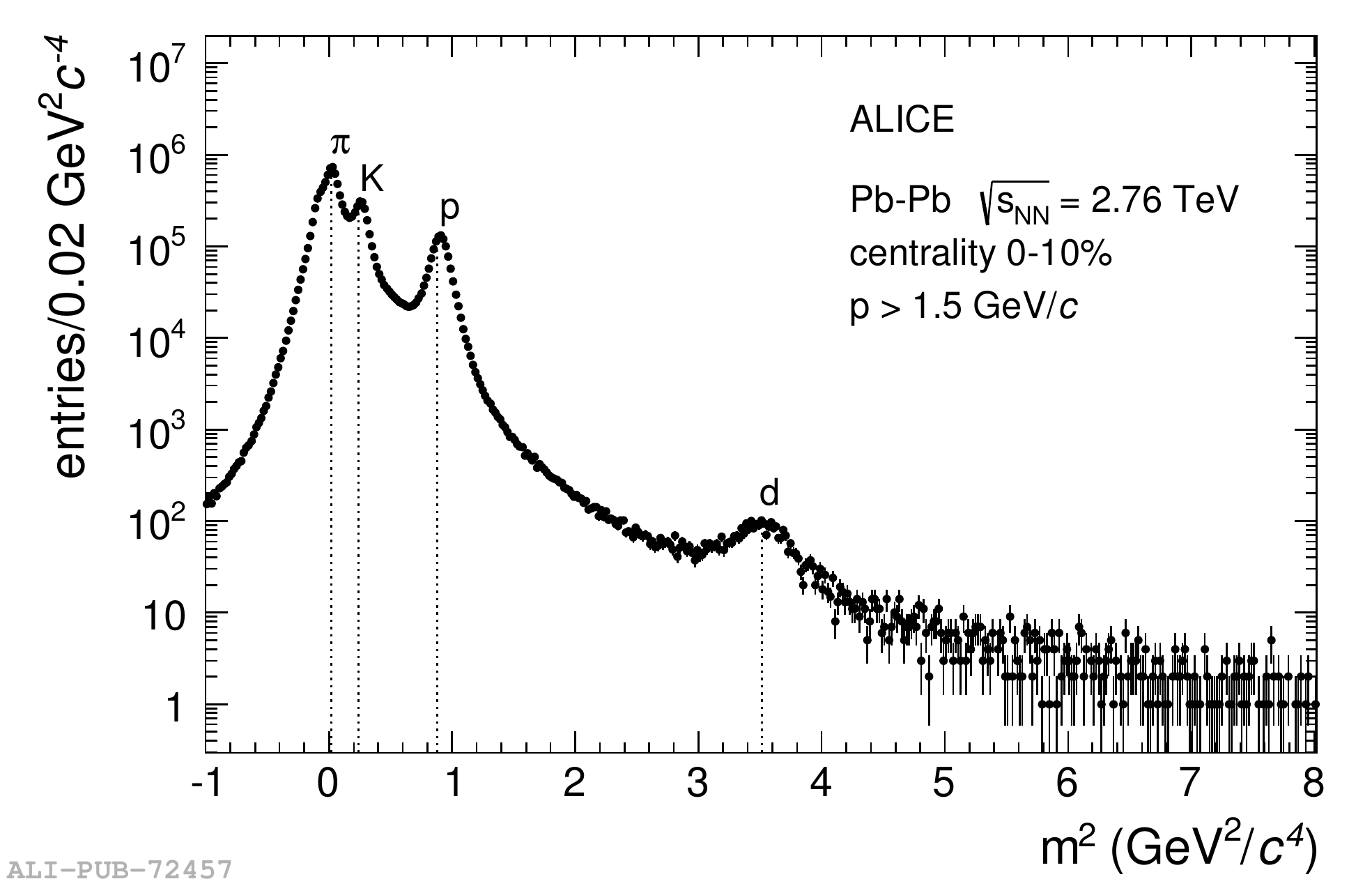}
\caption{Left: specific energy loss (d$E$/d$x$) in the TPC vs. rigidity in Pb--Pb collisions at $\sqrt{s_{\rm{NN}}}$~=~2.76~TeV. The lines show the parametrization of the expected mean energy loss. Right: squared particle masses calculated from the momentum and Cherenkov angle 
reconstructed with ITS-TPC and HMPID, respectively, in central Pb--Pb collisions at $\sqrt{s_{\rm{NN}}}$~=~2.76~TeV. 
Dotted lines indicate the PDG mass values. The pion tail on the left-hand side is suppressed by an upper cut on the Cherenkov angle. The deuteron peak is clearly visible~\cite{ALICE_PERFORMANCE}.}
\label{PERF}       
\end{figure}

\begin{figure}
\centering
\includegraphics[width=6.2cm,clip]{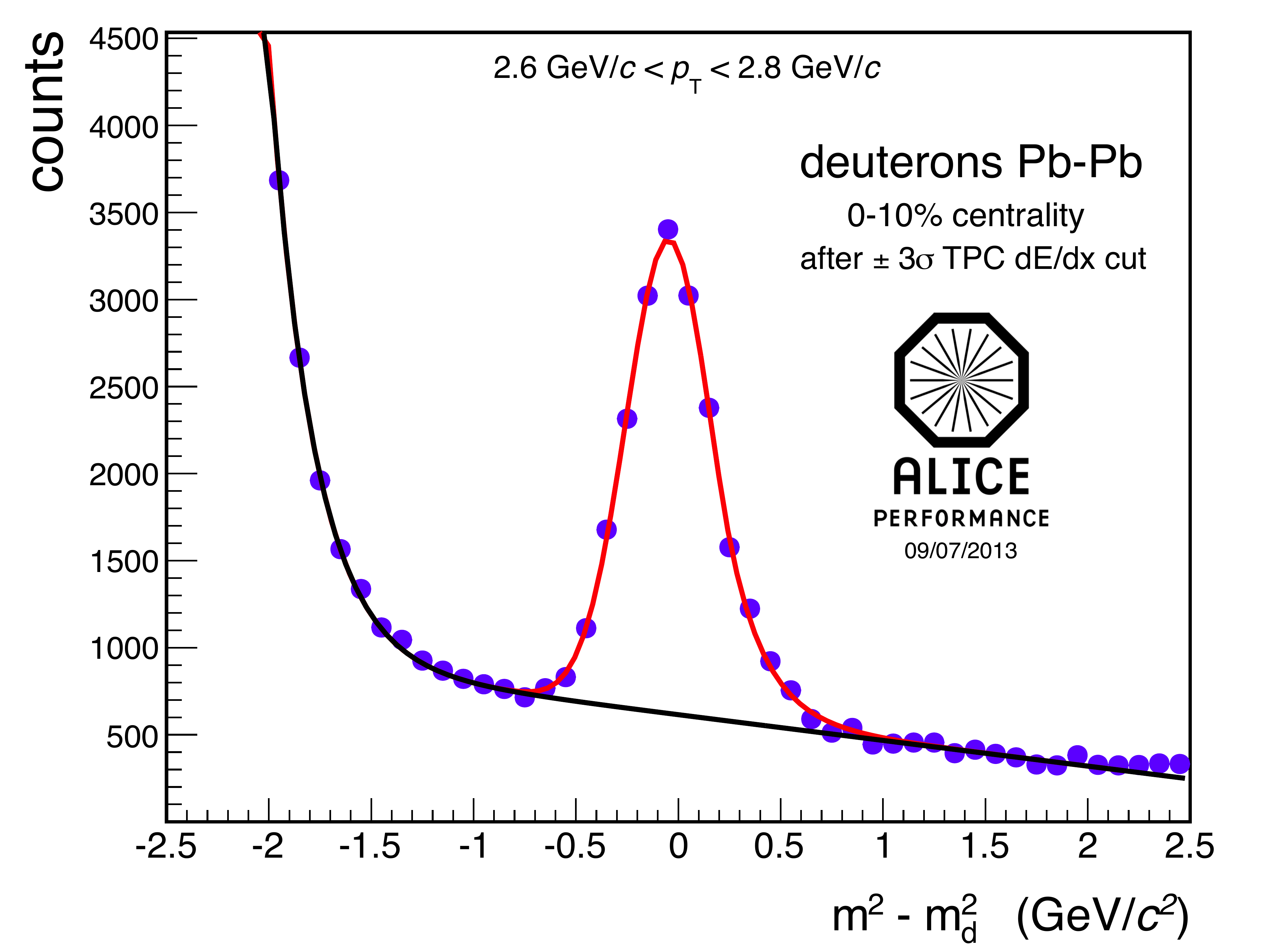}
\includegraphics[width=6.2cm,clip]{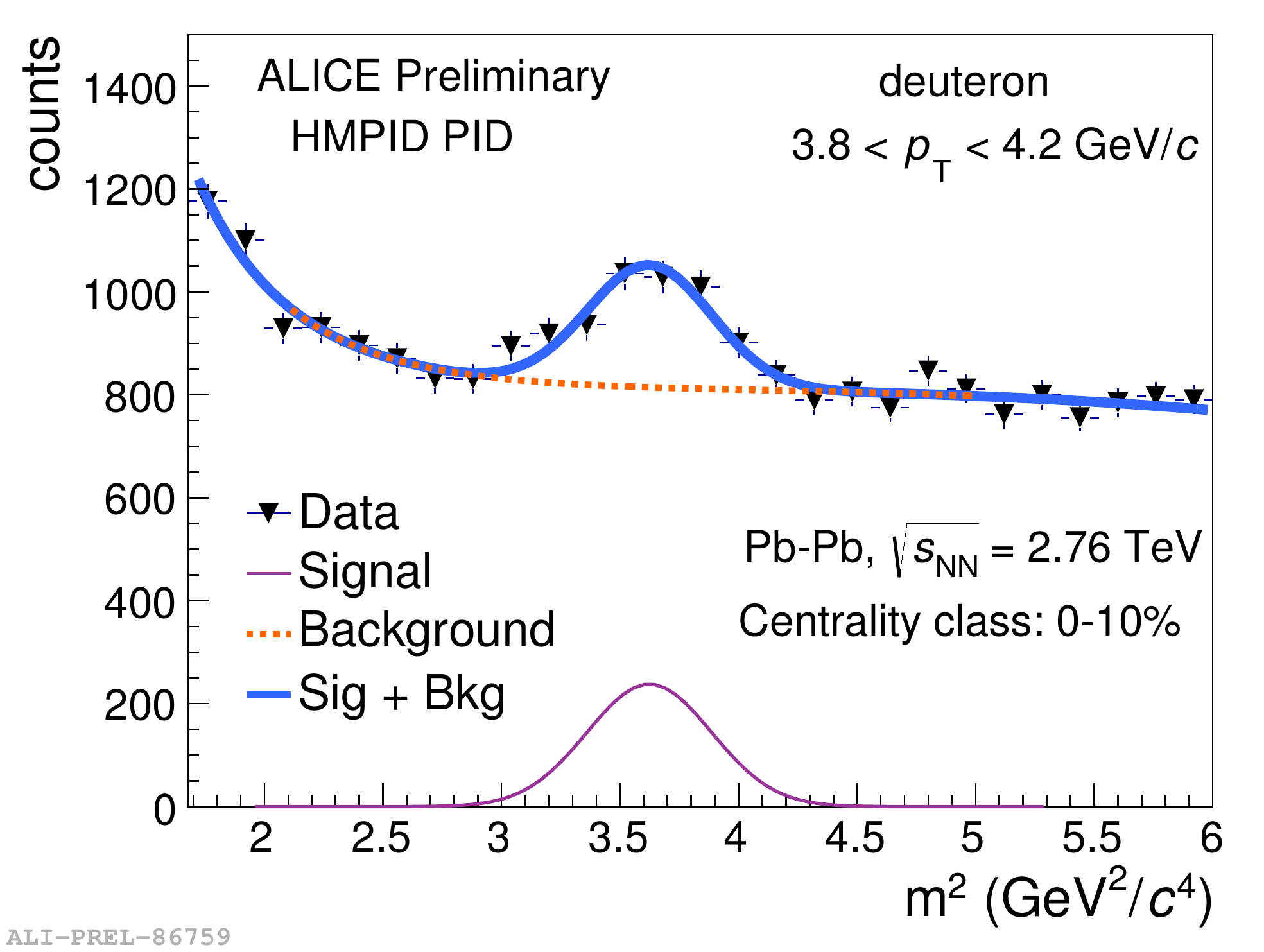}
\caption{Left: distribution of ($m^{2}_{\rm{TOF}}$ -- $m^{2}_{\rm{d}}$)  measured with the TOF detector (after a 3$\sigma$ TPC d$E$/d$x$ cut) for positive tracks with 2.6~<~$\textit{p}_{\rm{T}}$~<~2.8~GeV/$c$ in Pb--Pb collisions at $\sqrt{s_{\rm{NN}}}$~=~2.76~TeV (0-10$\%$ centrality class). The background from mismatched tracks (black solid line) is subtracted to obtain the raw yields. Right: distribution of m$^{2}$ measured with the HMPID detector for positive tracks with 3.8~<~$\textit{p}_{\rm{T}}$~<~4.2~GeV/$c$ in Pb--Pb collisions at $\sqrt{s_{\rm{NN}}}$~=~2.76~TeV (0-10$\%$ centrality class). The background (orange dotted line) is subtracted to obtain the raw yields. The peak corresponding to the deuteron mass is clearly visible for both distributions.}
\label{EXAM}       
\end{figure}

\begin{figure}
\centering
\includegraphics[width=8cm,clip]{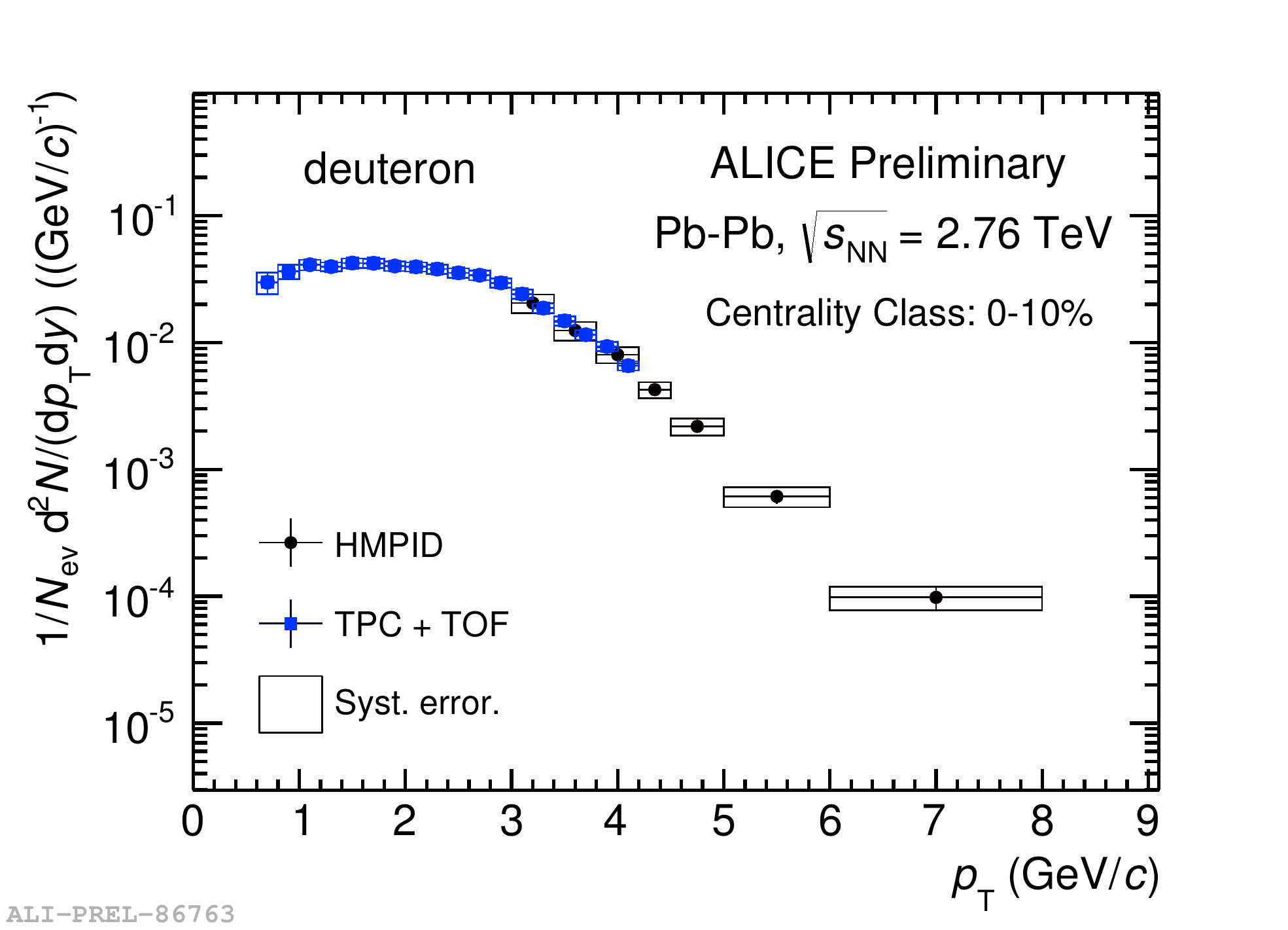}
\caption{Efficiency and acceptance corrected deuteron spectra using the TPC-TOF analysis (blue marker) and HMPID (black marker) in Pb--Pb collisions at $\sqrt{s_{\rm{NN}}}$~=~2.76~TeV in 0-10$\%$ centrality class. The agreement between the two different PID techniques, in the common $\textit{p}_{\rm{T}}$ interval (3~<~$\textit{p}_{\rm{T}}$~<~4.2~GeV/$c$), is within the total systematic uncertainties.
}
\label{COMP}       
\end{figure}

\begin{figure}
\centering
\includegraphics[width=7.2cm,clip]{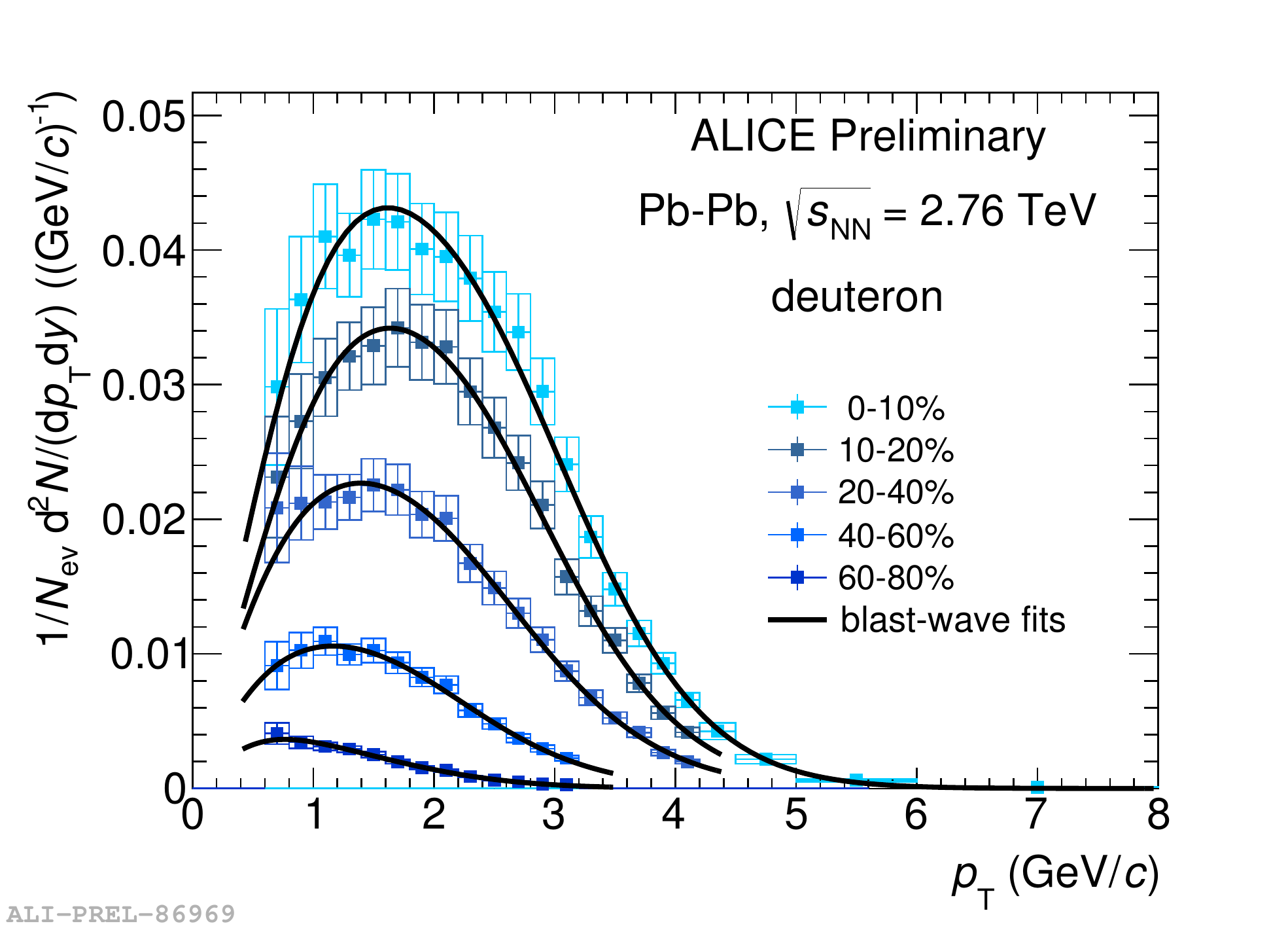}
\includegraphics[width=6cm,clip]{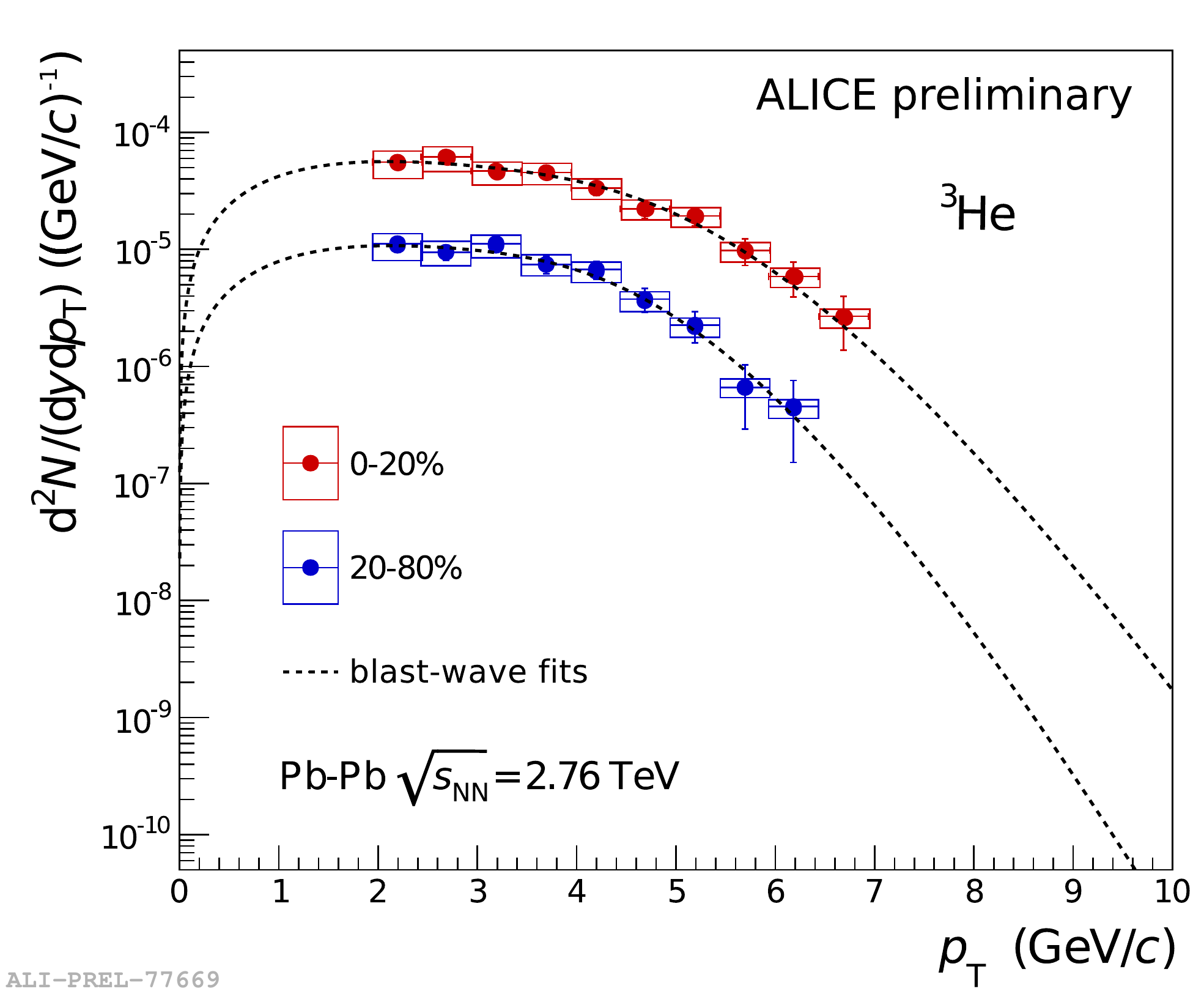}
\caption{Left: efficiency and acceptance corrected deuteron spectra for Pb--Pb collisions at $\sqrt{\rm{s}_{NN}}$ = 2.76 TeV in various centrality classes. The black lines represent an individual fit with the blast wave function. The boxes show the systematic error while the vertical lines represent the statistical error.
Right: ${}^3\rm{He}$ spectra for Pb--Pb collisions. The spectra are shown for two centrality classes (0-20$\%$ and 20-80$\%$) and fitted individually with the blast wave function. The systematic and statistical uncertainties are shown by boxes and vertical lines, respectively.
}
\label{spectra}       
\end{figure}

\begin{figure}
\centering
\includegraphics[width=8cm,clip]{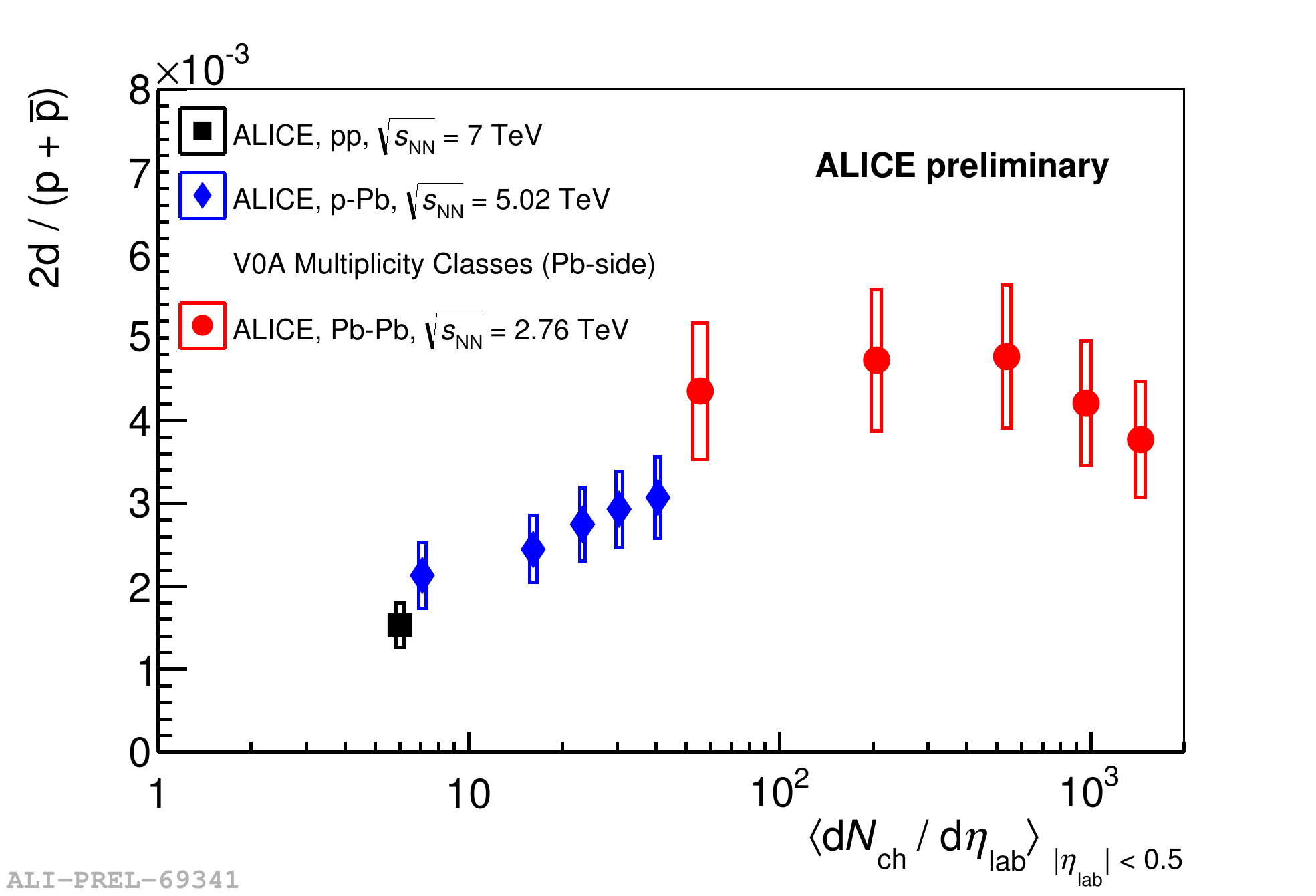}
\caption{Deuteron-to-proton ratio as a function of charged-particle multiplicity at midrapidity.}
\label{ratiogen}       
\end{figure}

Nuclei and anti-nuclei such as (anti-)deuterons, (anti-)tritons, and (anti-)${}^3\rm{He}$ are identified using the specific energy loss measurement in the TPC as already shown in Figure~\ref{PERF} (see Section~\ref{Sec2}). The dashed curves represent a parametrization of the Bethe-Block function for the different particle species. The measured energy-loss signal of a track is required to be within 3$\sigma$ from the expected value for a given mass hypothesis. This method provides a pure sample of ${}^3\rm{He}$ nuclei in the relevant $\textit{p}_{\rm{T}}$-range between 2~GeV/$c$ and 8~GeV/$c$ and deuterons up to 1.4~GeV/$c$.\\
 Figure~\ref{EXAM} (left) shows the ($m^{\rm{2}}$ - $m^{\rm{2}}_{\rm{d}}$) distribution for 2.6~<~$\textit{p}_{\rm{T}}$~<~2.8~GeV/$c$ obtained by matching of the TOF and TPC information. The signal is on top of a significant background originating from the association of a track with a TOF signal and the non-gaussian tails of lower mass particles. For each $\textit{p}_{\rm{T}}$-interval, this $m^{2}$-distribution is fitted with a Gaussian function with an exponential tail for the signal and an exponential function added to a first order polynomial for the background.\\  
Deuterons and anti-deuterons are also identified at high-$\textit{p}_{\rm{T}}$, namely in the range  3~<~$\textit{p}_{\rm{T}}$~<~8~GeV/$c$,
in central Pb--Pb collisions by using the HMPID detector. 
The analysis in this case is limited by the available statistics 
and therefore the (anti-)deuteron measurements are performed only in the 0-10$\%$ centrality class. 
The $m^{\rm{2}}$-distributions are derived from the $\rm{cos}(\theta_{\rm{cherenkov}})$~=~$1/(n \beta)$ and the momentum information from the tracking systems:
\begin{eqnarray}
m^2  = p^{2} (n^{2} \rm{cos^{2}} \theta_{\rm{cherenkov}} -1) 
\end{eqnarray}
Figure~\ref{EXAM} (right) shows the distribution of $m^{2}$-measured with the HMPID detector for positive tracks with 3.8~<~$\textit{p}_{\rm{T}}$~<~4.2~GeV/$c$ in Pb--Pb collisions at $\sqrt{s_{\rm{NN}}}$~=~2.76 TeV (0-10$\%$ centrality class). In central Pb--Pb collisions, where the total number of hits in the HMPID is large, a Cherenkov ring could be reconstructed with hits incorrectly associated to the track, thus producing the significant background under the deuteron signal~\cite{highptpaper}. 
For each $\textit{p}_{\rm{T}}$-interval, the $m^{2}$-distribution is fitted with a Gaussian function for the signal and a negative power law function added to a third order polynomial for the background. 
Figure~\ref{COMP} shows the comparison between the efficiency and acceptance corrected deuteron spectra measured with the TPC-TOF and HMPID for the 10$\%$ most central Pb--Pb collisions at $\sqrt{s_{\rm{NN}}}$~=~2.76~TeV. The agreement between the two different PID techniques, in the common $\textit{p}_{\rm{T}}$ interval (3~<~$\textit{p}_{\rm{T}}$~<~4.2~GeV/$c$) is within the total systematic uncertainties.\\
Figure~\ref{spectra} shows efficiency and acceptance corrected spectra of d and ${}^3\rm{He}$, obtained in Pb--Pb collisions for different centrality selections.  
The $\textit{p}_{\rm{T}}$ distributions show a clear evolution, becoming harder as the centrality increases, a behaviour similar to that of the protons which exhibit a significant radial flow~\cite{ALICEPbPb}. The spectra 
obtained in Pb--Pb collisions are individually fitted with the blast wave model for the determination of $\textit{p}_{\rm{T}}$-integrated yields. Figure~\ref{ratiogen} shows the deuteron-to-proton ratio as a function of the multiplicity in pp, p--Pb and in Pb--Pb collisions. The ratio rises with multiplicity until a possible saturation is reached for Pb--Pb collisions.


\begin{figure}
\centering
\includegraphics[width=6.5cm,clip]{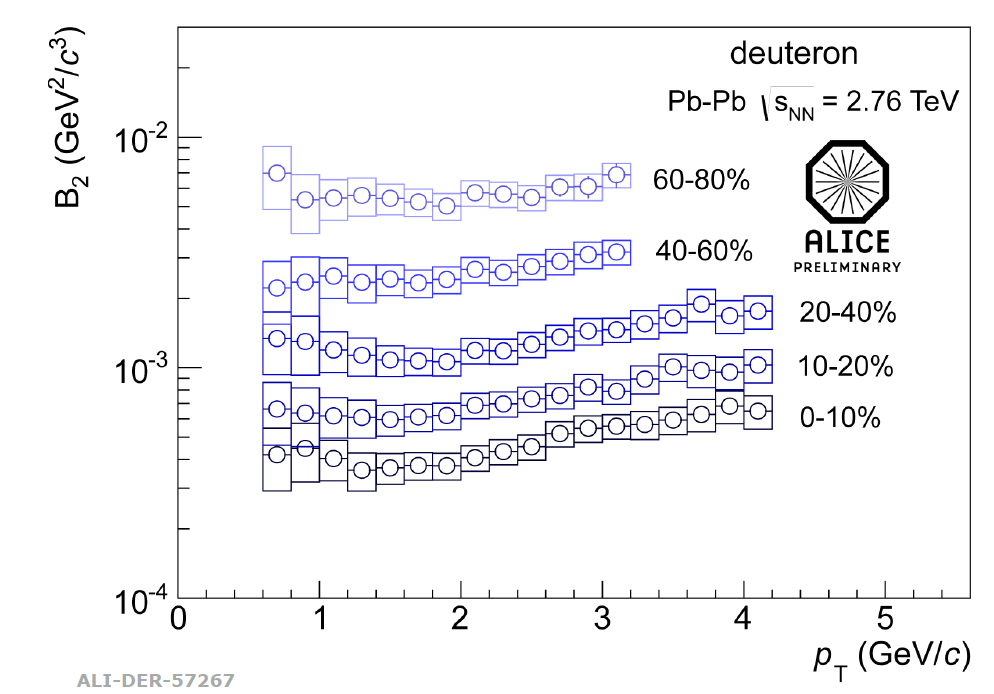}
\includegraphics[width=7.cm,clip]{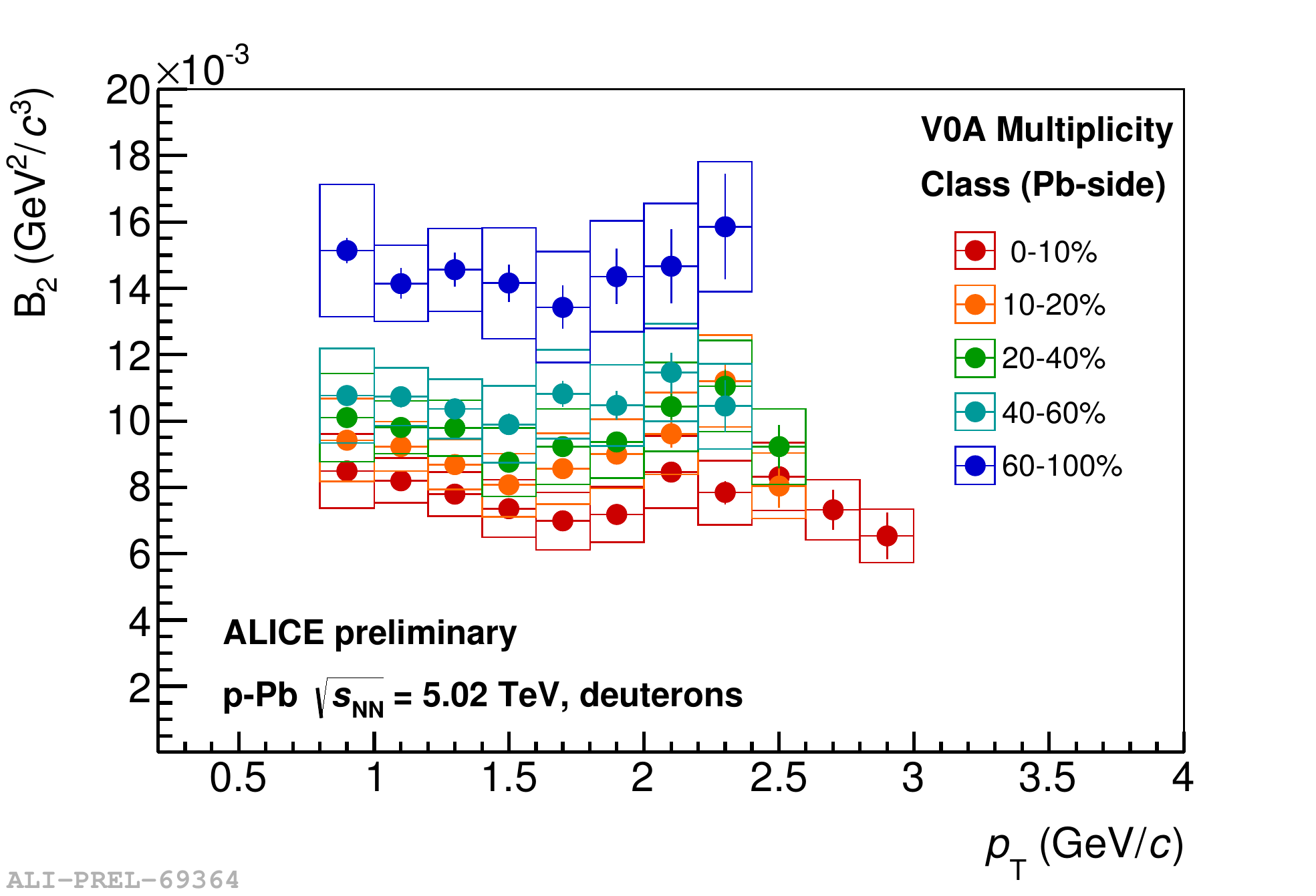}
\caption{Coalescence parameter $B_{\rm{2}}$ as a function of the transverse momentum per nucleon in Pb--Pb collisions at $\sqrt{\rm{s}_{NN}}$~=~2.76~TeV (left) and in p--Pb collisions at $\sqrt{\rm{s}_{NN}}$~=~5.02~TeV (right).}
\label{B2}       
\end{figure}
 In the coalescence approach, light nuclei are formed after the kinetic freeze-out via coalescence of protons and neutrons which are near in space and with a low relative momentum. In this mechanism, the spectral distribution of the composite nuclei is related to the one of the primordial nucleons via the following relationship:
\begin{eqnarray}
E_{i} \frac{d^{3}N_i}{(dp_i)^3} = B_A \bigg(E_p \dfrac{d^3 N_p}{(dp_p)^3}\bigg)^A
\end{eqnarray}
assuming that protons and neutrons have the same momentum distribution. $B_{\rm{A}}$ is the coalescence parameter of particle $i$ with mass number $A$ and a momentum of $p_{\rm{i}}$ = Ap$_{p}$. \\
Figure~\ref{B2} shows the obtained $B_{\rm{2}}$ values for deuterons for Pb--Pb (the measurement is currently being extended to higher transverse momentum using the HMPID) and p--Pb (right) collisions. A clear decrease with increasing centrality is observed for Pb--Pb collisions. In the coalescence picture, this behaviour is explained by an increase in the source volume V$_{eff}$. $B_{\rm{2}}$ also shows an increasing trend with the transverse momentum for central collisions, in contrast with the most simple coalescence models. This behaviour can be qualitatively understood by position-momentum correlations which are caused by a radially expanding source. In p--Pb collisions the $B_{\rm{2}}$ is slightly decreasing with multiplicity, again in contrast with the most simple coalescence models.

\begin{figure}
\centering
\includegraphics[width=6.2cm,clip]{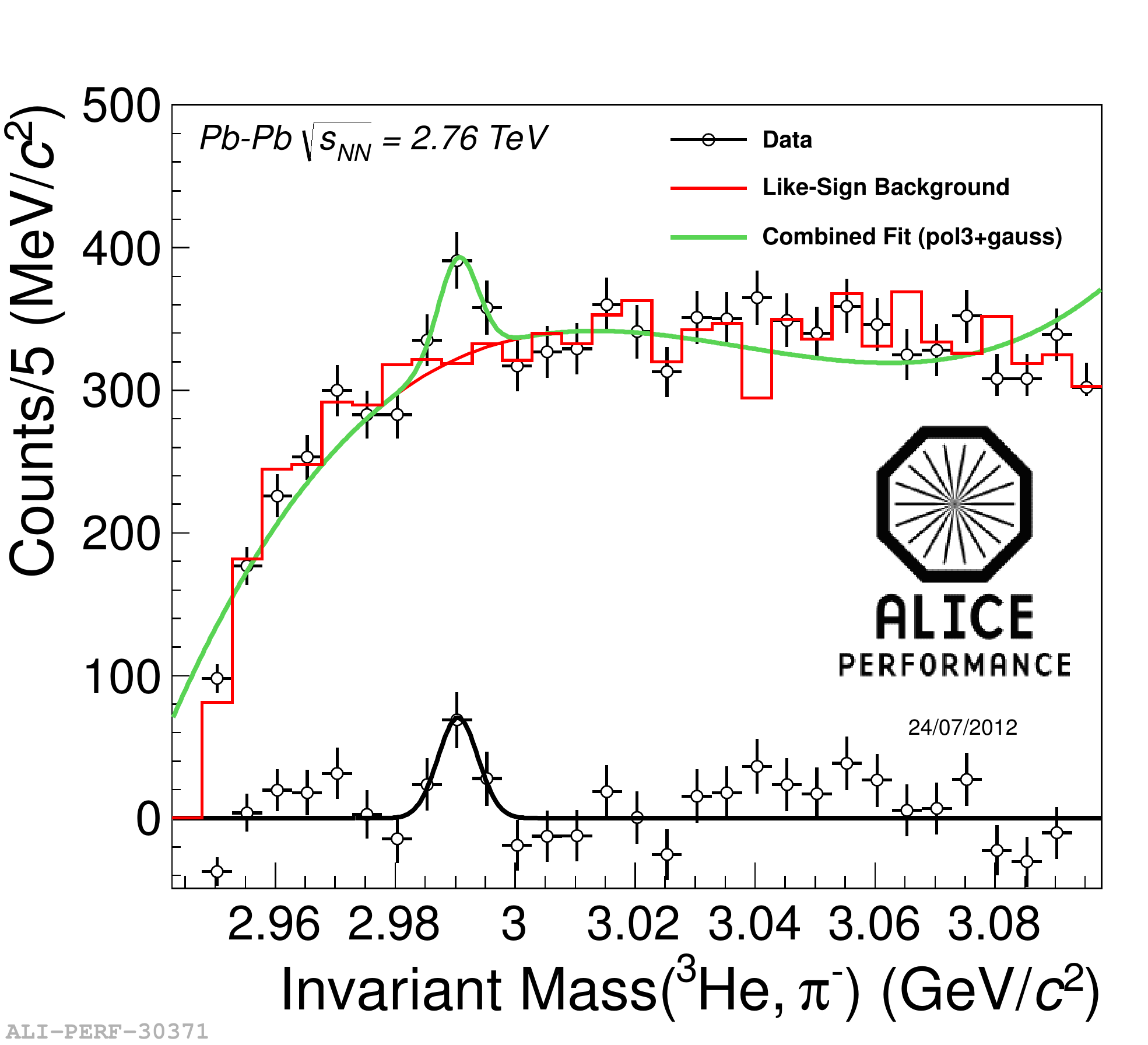}
\includegraphics[width=6.cm,clip]{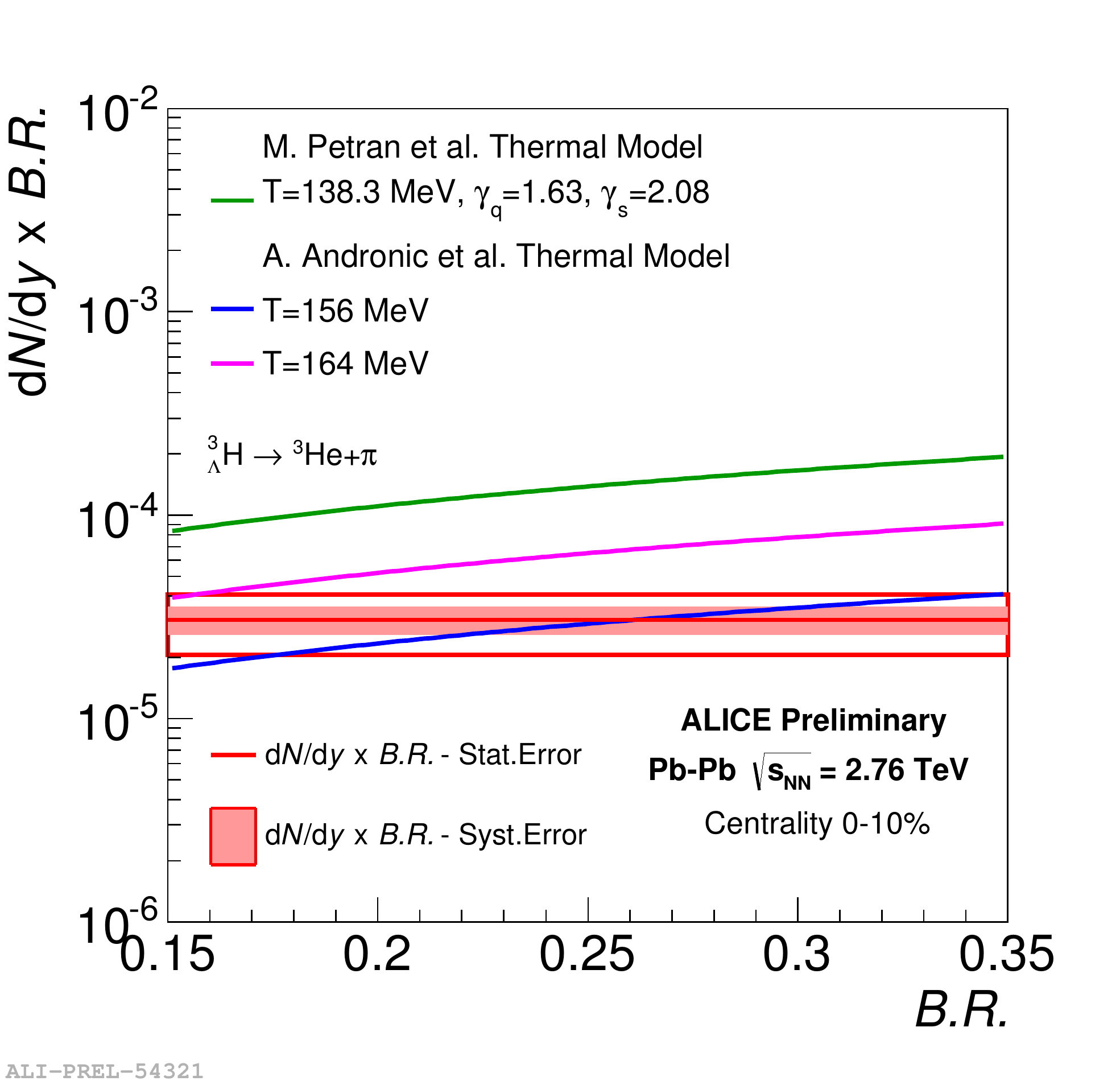}
\caption{Left: invariant mass of (${}^3\rm{He}$, $\pi^{-}$) for Pb--Pb collisions. Right: d$N$/d$y$ in comparison with different models for the hypertriton measurement.}
\label{HYP}    
\end{figure}

\section{Hypertriton}

The hypertriton ${}^3_\Lambda{\rm{H}}$ is the lightest known hypernucleus formed by a proton, a neutron and a $\Lambda$. The production of hypertriton ${}^3_\Lambda{\rm{H}}$ has been measured in Pb--Pb collisions via invariant mass reconstruction in the decay channel ${}^3_\Lambda{\rm{H}}$ $\rightarrow$  ${}^3{\rm{He}}$ + $\pi^{-}$. The hypertriton has a mass of 2.991 $\pm$ 0.002 GeV/$c^{\rm{2}}$ and a decay length on the order of the free $\Lambda$ particle. The daughter tracks of the hypertriton candidates are required to originate from a secondary vertex and identified as a ${}^3{\rm{He}}$ and a pion via TPC d$E$/d$x$. The resulting invariant mass distribution is shown in figure~\ref{HYP} (left). As the production yield d$N$/d$y$ depends on the branching ratio (B.R.), the d$N$/d$y$ is computed to different models as a function of the B.R. in figure~\ref{HYP} (right). For the most likely B.R. of 25$\%$, the measured d$N$/d$y$ agrees very well with the equilibrium thermal model prediction for a temperature value of 156 MeV.

\section{Exotica bound states}

\begin{figure}
\centering
\includegraphics[width=7.cm,clip]{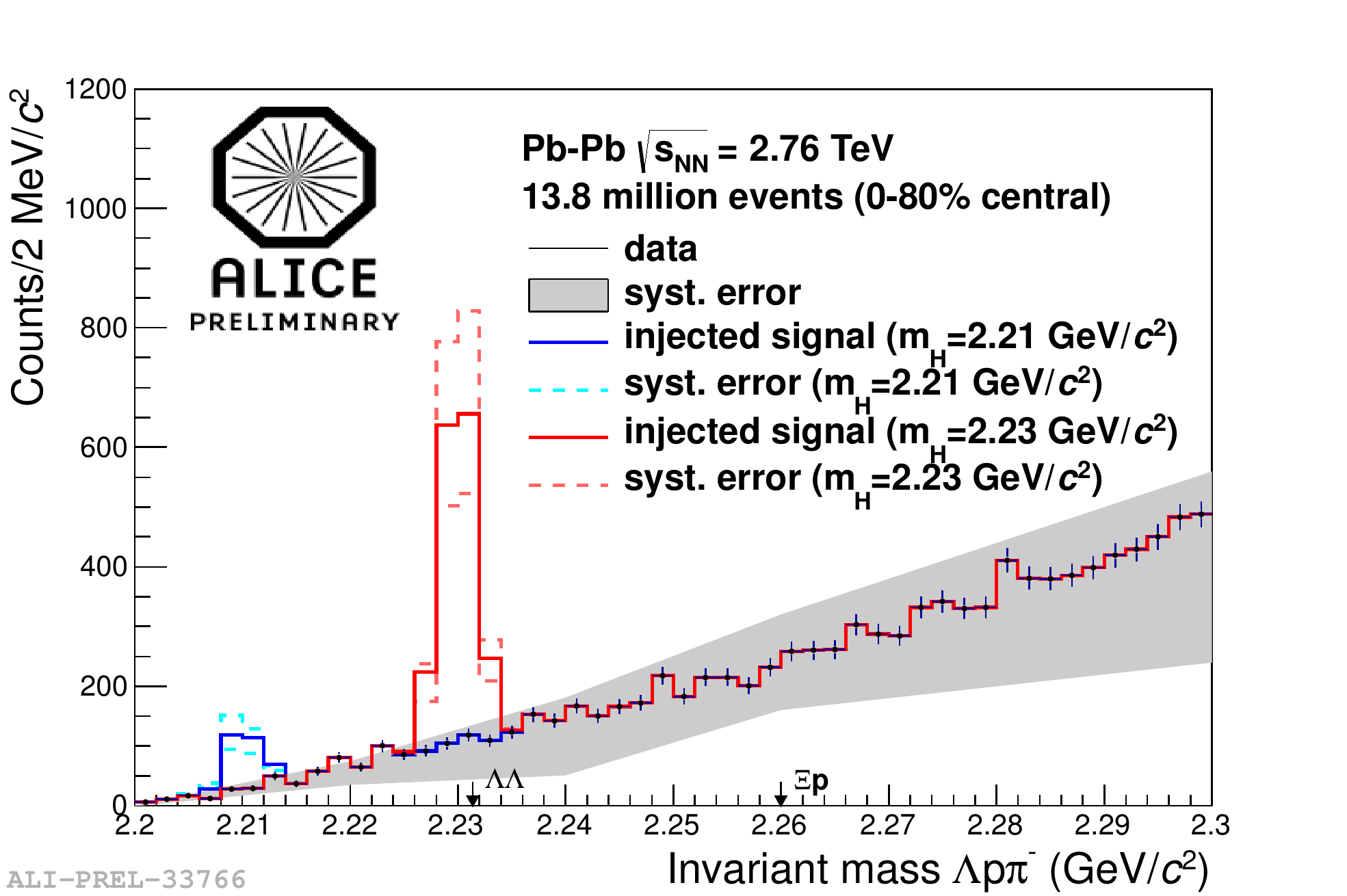}
\includegraphics[width=6.cm,clip]{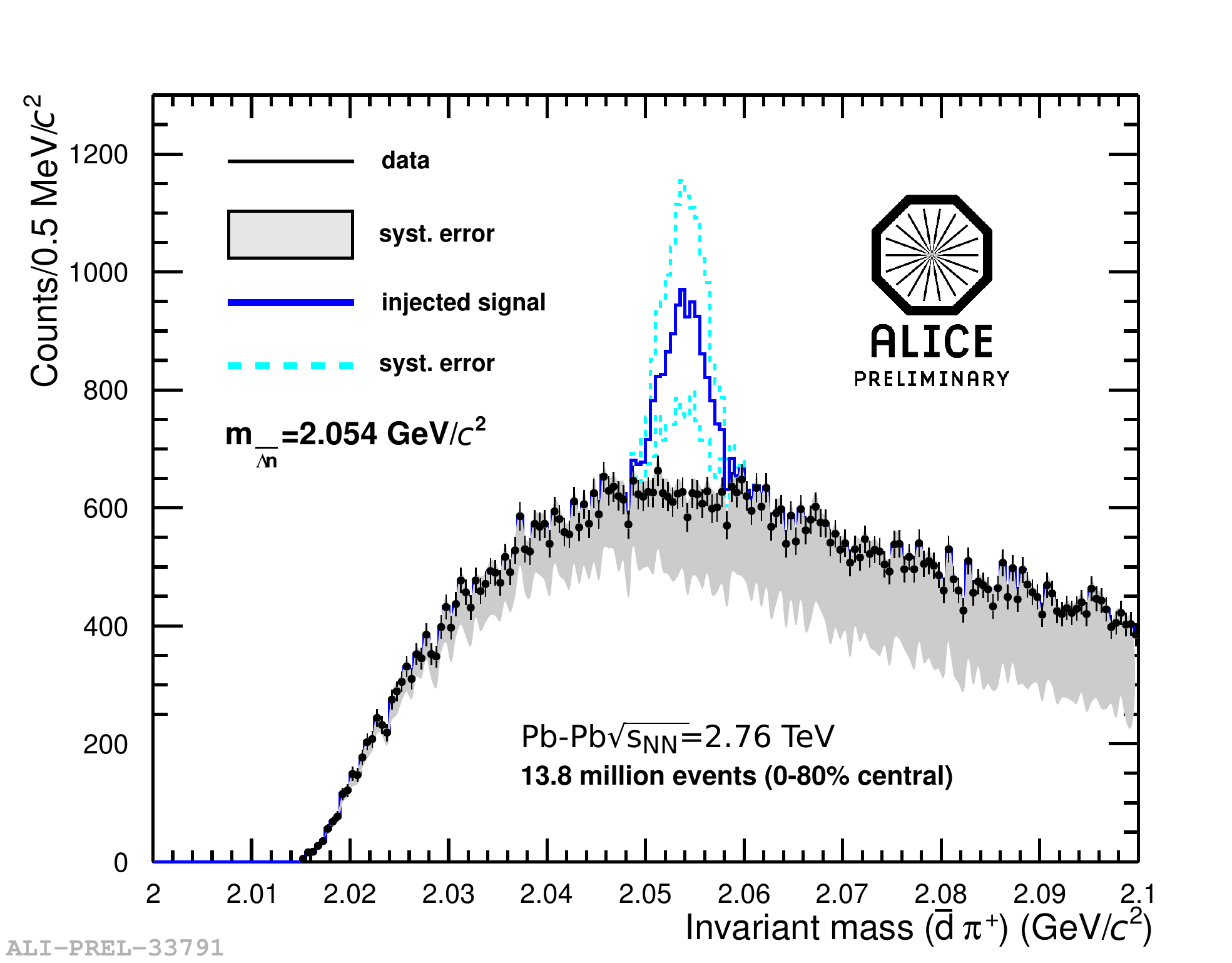}
\caption{Left: invariant mass of H-dibaryon with injected signal for a weakly bound H in red (m$_{\rm{d}}$~=~2.23~GeV/$c^{\rm{2}}$) and a stronger bound H in blue (m$_{\rm{H}}$~=~2.21~GeV/$c^{\rm{2}}$). The arrow at 2.231~GeV/$c^{\rm{2}}$ indicates the $\Lambda$$\Lambda$ threshold and the arrow at 2.260~GeV/$c^{\rm{2}}$ the $\Xi$p threshold. The systematic errors are shown for data in grey and for the injected signal as dashed lines. Right: invariant mass of the anti-deuteron and negative pion with injected signal (blue line).
}
\label{MASSHDIB}     
\end{figure}

Searches for the H-dibaryon and the $\Lambda$n bound state were performed in Pb--Pb collisions from the 2010 data taking (13.8 $\times$ 10$^6$ minimum bias events). The H-dibaryon is a bound state of $uuddss$ ($\Lambda\Lambda$) first predicted by Jaffe~\cite{EXOT0} in a bag model calculation. Recent lattice QCD calculations\footnote{These calculations have been performed at an unphysical pion mass of about 390 GeV/$c$$^{\rm{2}}$.} ~\cite{EXOT1, EXOT2} also suggest a bound state with binding energies in the range 13-50 MeV. A chiral extrapolation of these lattice calculations to a physical pion mass resulted in a H-dibaryon unbound by either 13~$\pm$~14~MeV or close to the $\Xi$p threshold. The possible existence of the weakly bound H-dibaryon has been investigated via the decay channel H $\rightarrow$ $\Lambda$ + p + $\pi$.\\ 
Figure~\ref{MASSHDIB} (left) shows the invariant mass of $\Lambda$ + p + $\pi$. No evidence of a signal for the H-dibaryon has been found with the currently available statistics. The expected signal for two different possible bound states of the H-dibaryon are also shown. This signal was computed estimating the acceptance $\times$ efficiency (from Monte-Carlo simulation), the production rates as predicted by the thermal model~\cite{EXOT3} and the expected branching ratios~\cite{EXOT4}. In the Monte Carlo simulation the lifetime of the H-dibaryon has been assumed to be the same as the free $\Lambda$. The upper limit for the production yield of d$N$/d$y$~<~2~$\cdot$~10$^{-4}$ (99$\%$CL) is obtained for a 1 MeV bound H-dibaryon. 
Also in the invariant mass distribution of deuterons and pions from displaced vertices, where a possible $\Lambda$n bound state is expected to be visible, no signal can be seen (Figure~\ref{MASSHDIB}, right). This corresponds to an upper limit of d$N$/d$y$~<~1.5~$\cdot$~10$^3$ (99$\%$CL). The obtained upper limits for H-dibaryon and $\Lambda$n bound state are more than a factor 10 below the different model predictions~\cite{PRED1,PRED2} . 

\end{document}